\documentclass[authoryear,preprint,review,12pt]{elsarticle}
\usepackage{amsmath}
\usepackage{amssymb}
\usepackage{graphicx}
\usepackage{float}
\usepackage{placeins}
\usepackage{amsthm}
\usepackage{enumerate}
\usepackage{verbatim}
\usepackage{multirow}
\usepackage[margin=1in]{geometry}

\usepackage[round]{natbib}

\usepackage{booktabs}

\usepackage{makecell}

\usepackage[mathlines]{lineno}

\allowdisplaybreaks

\begin{document}

\title{Improved confidence intervals for nonlinear mixed-effects and nonparametric regression models%\thanks{Grants or other notes
}
%\subtitle{Improved CI for nonlinear mixed-effects models}


\author[]{Nan Zheng$^a$, Noel Cadigan$^b$\corref{corr}}	
%\author[]{Noel Cadigan$^b$\corref{corr}}

\address{$^a$Department of Mathematics and Statistics, \\ Memorial University of Newfoundland, St. John's, NL, Canada. A1C 5S7\\\vspace{3mm}
$^b$Centre for Fisheries Ecosystems Research,\\ Fisheries and Marine Institute of Memorial University of Newfoundland, St. John’s, NL, Canada. A1C 5R3
}

\cortext[corr]{Corresponding author. E-mail address: noel.cadigan@mi.mun.ca}

\maketitle

\begin{abstract}
Statistical inference for high dimensional parameters (HDPs) can be based on their intrinsic correlation; that is, parameters that are close spatially or temporally tend to have more similar values. This is why nonlinear mixed-effects models (NMMs) are commonly (and appropriately) used for models with HDPs. Conversely, in many practical applications of NMM, the random effects (REs) are actually correlated HDPs that should remain constant during repeated sampling for frequentist inference. In both scenarios, the inference should be conditional on REs, instead of marginal inference by integrating out REs. In this paper, we first summarize recent theory of conditional inference for NMM, and then propose a bias-corrected RE predictor and confidence interval (CI). We also extend this methodology to accommodate the case where some REs are not associated with data. Simulation studies indicate that this new approach leads to substantial improvement in the conditional coverage rate of RE CIs, including CIs for smooth functions in generalized additive models, as compared to the existing method based on marginal inference.
\end{abstract}


\section{Introduction}\label{sec:Intro}
The joint density function of a mixed-effects model (linear and nonlinear) can be written, with wide generality, as $f(D,\Psi|\theta)=f(D|\Psi,\theta)f(\Psi|\theta)$, where $D, \Psi$ and $\theta$ are respectively data, random effects (REs) and model parameters (FEs, fixed effects), and $f(\cdot)$ represents the corresponding probability density/mass function (pdf/pmf).
The logarithm of the joint likelihood of data and REs is
\begin{linenomath*} 
	\begin{align}\label{EQ:lj_November_28_2022}
		\begin{split}
			l_j(\Psi,\theta) = \log f(D,\Psi|\theta) &= \log f(D|\Psi,\theta) + \log f(\Psi|\theta)
			= l_c + l_r,			
		\end{split}	 
	\end{align}
\end{linenomath*} 
where $l_c=\log f(D|\Psi,\theta)$ and $l_r=\log f(\Psi|\theta)$ denote respectively the log-likelihood conditional on REs and the log-likelihood of REs. We are addressing a general class of models with broad applicability in the ecology and evolution field. These models, while not limited to, include integrated state-space fishery assessment models that incorporate various data sources depending on the fish stock. We do not introduce notation for specific models and data, but see \cite{aeberhard2018review} for an example. We do provide simpler examples of random walk state-space models and a generalized additive model.

In many applications of mixed-effects models, the objective of statistical inference is for a single set of REs, which are considered to be fixed and frequentist inference is based on repeated sampling of the data given the REs.
Basically, the REs are conceptually treated as FEs, possibly with high dimension, and they are modelled as REs with marginal distribution $f(\Psi|\theta)$ for pragmatic purposes.
This is opposed to basing statistical inference on repeated sampling of the REs and the data, which is the common statistical basis assumed for mixed-effects models. 
For example, in a fisheries stock assessment model, even though the population abundances for consecutive years may be modeled as REs following a yearly AR(1) (auto-regressive model of order 1) model, the population abundance for a specific year remains fixed for sampling events in different locations and dates of the year. 
In this context, statistical inference should be conditional on the true but unknown REs, instead of integrating out REs over their marginal distribution $f(\Psi|\theta)$. We refer to the former inferential setting as ``conditional'', and the latter as ``marginal''.

Frequently, a substantial part of the domain of $f(\Psi|\theta)\gg0$ involves values of $\Psi$ that can produce unrealistic data via $f(D|\Psi,\theta)$, such as the extinction of a fish stock, in contradiction to the observed data.
Nevertheless, in marginal inference, the properties of RE predictors $\hat{\Psi}(\hat{\theta})$ and parameter estimators $\hat{\theta}$, such as the mean $\mathrm{E}\{\hat{\Psi}(\hat{\theta})\}$ and prediction error covariance $\mathrm{Cov}\{\hat{\Psi}(\hat{\theta})-\Psi\}$, are evaluated by integrating over all values of the REs (realistic and otherwise) and the resulting sample space, leading to biased conclusion.
For instance, if $\hat{\Psi}(\hat{\theta})$ is the posterior mode and the posterior distribution of REs is approximately symmetric, then the marginal bias $\mathrm{E}\{\hat{\Psi}(\hat{\theta})-\Psi\}=O(1/T)$ \citep{zheng2021frequentist}, while the conditional bias $\mathrm{E}\{\hat{\Psi}(\hat{\theta})-\Psi|\Psi\}=O(1)$ \citep{Zheng2021P2}.
In this paper $T$ denotes the total number of observational units \citep[e.g.,][]{kass1989approximate, flores2019bootstrap}.
For example, in a yearly time-series setting, $T$ indicates the number of years.
The sample size for each observational unit substantially affects the accuracy and precision of RE predictions.
However, such unit-wise sample sizes can vary from zero to large numbers. In order to accommodate all these cases, we do not impose restriction on the unit-wise sample sizes and hence we do not develop notations for them.

With parametric empirical Bayes inference \citep{kass1989approximate}, the posterior mean and variance of $\Psi$ given the observed data $D$ are used respectively as the measures of central tendency and variability of the RE predictions, and the results of \cite{kass1989approximate} are widely adopted by modeling packages for nonlinear mixed-effects models, such as Template Model Builder \citep[TMB,][]{kristensen2015tmb} and AD Model Builder \citep[ADMB,][]{fournier2012ad}, to construct prediction intervals for REs and nonlinear functions of REs.
However, the posterior mean and variance in Bayesian inference are equal respectively to the frequentist marginal $\mathrm{E}\{\hat{\Psi}(\hat{\theta})\}$ and $\mathrm{Cov}\{\hat{\Psi}(\hat{\theta})-\Psi\}$ to the first order approximation \citep[proof in][]{zheng2021frequentist}; that is, inference given the data still results in marginal inference, and fundamentally this cannot provide accurate measures of variability of RE predictions when REs behave like high-dimensional FEs (namely, the conditional inferential setting).
Note that the covariance of the prediction error, $\mathrm{Cov}\{\hat{\Psi}(\hat{\theta})-\Psi\}$, is equal to the marginal mean squared error (MSE) of $\hat{\Psi}(\hat{\theta})$ to a first order approximation \citep{zheng2021frequentist}.
The simulation studies in \cite{Zheng2021P2} demonstrated that under the conditional setting, confidence intervals (CIs) constructed using marginal root MSEs can lead to coverage probabilities substantially different from the nominal levels.

These marginal inference methods have also been used with nonparametric regression models, such as using Bayesian posterior variance, or equivalently, the frequentist marginal MSE, to construct CIs in generalized additive models \citep[GAMs, e.g.,][]{marra2012coverage,wood2020inference}.
Here nonparametric regression models refer generally to the statistical models that have parametric and nonparametric components, and both components can be of interest.
When marginal inference is applied in the conditional inferential setting, component-wise CI coverage probabilities may be unreliable, and the ``across-the-function'' coverage property has been used to validate the inferential performance \citep{marra2012coverage,wood2020inference}.
If we still use $\Psi$ to denote the high-dimensional parameters, the component-wise CI coverage probability is for each component $\Psi_i$ of $\Psi$, and the across-the-function coverage probability is averaged across all the $\Psi_i$'s. \cite{Zheng2021P2} found that if marginal MSEs are used for conditional inference, in some cases even the across-the-function coverage probability can deviate substantially from the nominal one.

The penalized log-likelihood for a GAM model can be written as \citep{wood2016smoothing}
\begin{linenomath*} 
	\begin{align}\label{EQ:penalized_log_likelihood_GAM_November_28_2022}
		\mathcal{L}(\pmb{\beta}) &= l(\pmb{\beta}) - \dfrac{1}{2}\sum_{j=1}^M\lambda_j \pmb{\beta}^{\top}\pmb{S}_j\pmb{\beta}, 
	\end{align}
\end{linenomath*} 
where the vector $\pmb{\beta}$ includes both the basis coefficients and other model parameters $\theta$, $l(\pmb{\beta})$ is the likelihood, $\pmb{\beta}^{\top}\pmb{S}_j\pmb{\beta}$ are the smoothing penalties with known sparse matrices $\pmb{S}_j$, and $\lambda_j$ are the smoothing parameters. \cite{wood2016smoothing} suggested to estimate the model coefficients $\pmb{\beta}$ with the modes of $\mathcal{L}(\pmb{\beta})$, and the smoothing parameters $\lambda_j$ by maximizing the approximate marginal likelihood where $\pmb{\beta}$ are integrated out using the Laplace approximation.
Similar to $\lambda_j$, the $\theta$ parameters can also be estimated by maximizing the Laplace approximate marginal likelihood (LAML), where the basis coefficients are integrated out using the Laplace approximation.
This is equivalent to estimating both $\lambda_j$ and $\theta$ using REML (restricted maximum likelihood) in the sense of \cite{laird1982random}.
For this approach, we can denote the basis coefficients as $\Psi$, and re-write the penalized log-likelihood (\ref{EQ:penalized_log_likelihood_GAM_November_28_2022}) as
\begin{linenomath*} 
	\begin{align}\label{EQ:penalized_log_likelihood_GAM_LAML_November_28_2022}
		\begin{split}
			\mathcal{L}(\Psi,\theta) &= l(\Psi,\theta) - \dfrac{1}{2}\sum_{j=1}^M\lambda_j \Psi^{\top}\pmb{S}_j\Psi\\
			&= 	l(\Psi,\theta) - \dfrac{1}{2}\Psi^{\top}\left\lbrace \sum_{j=1}^M\lambda_j \pmb{S}_j\right\rbrace \Psi,		
		\end{split}	 
	\end{align}
\end{linenomath*} 
where the likelihood $l(\Psi,\theta)$ and the penalty term correspond respectively to $l_c$ and $l_r$ in (\ref{EQ:lj_November_28_2022}).

The GAM penalized log-likelihood (\ref{EQ:penalized_log_likelihood_GAM_LAML_November_28_2022}) and the mixed-effects model log-joint likelihood (\ref{EQ:lj_November_28_2022}) can be optimized in the same way with maximum LAML for $\theta$ and $\lambda_j$, and the posterior mode for $\Psi$.
The way the parameters $\theta$ are estimated is the only difference between GAM penalized log-likelihoods (\ref{EQ:penalized_log_likelihood_GAM_November_28_2022}) and (\ref{EQ:penalized_log_likelihood_GAM_LAML_November_28_2022}).
The correlation structure of $l_r$ in (\ref{EQ:lj_November_28_2022}) can involve some model parameters such as the autocorrelation parameter in an AR(1) model, while that of (\ref{EQ:penalized_log_likelihood_GAM_LAML_November_28_2022}) does not.
In this sense, GAM models can be regarded as special cases of mixed-effects models in the conditional setting \citep[see][]{wand2003smoothing,wood2013straightforward}.
Maximum LAML estimation of the model parameters results in a bias of $O(T^{-1/2})$ in the conditional setting \citep{Zheng2021P2}. Such bias generally appears in semi-parametric inference \citep[e.g.,][]{he2016flexible, cheng2018bias}, where in some cases the bias can be larger \citep[see, e.g., Eq. 79 in][]{zheng2018inferences}. Despite the biases, the variances of the maximum LAML estimators in the conditional setting are in general smaller than the corresponding values for the marginal setting due to the restriction on randomization when $\Psi$ is fixed \citep{Zheng2021P2}. Overall, the MSEs of the maximum LAML estimators in the conditional setting are equal to the marginal MSEs with an approximation order of $o(1/T)$ \citep{Zheng2021P2}.
In the marginal setting, the MSEs are the same as the variances of maximum marginal likelihood estimators (MMLEs), because the biases are negligible when averaged over the marginal distribution of $\Psi$.
Due to the consistency and efficiency of maximum likelihood estimators (MLEs), using maximum LAML estimators for $\theta$ and $\lambda_j$ in the conditional setting is well justified.
In this work, we will extend the methodology of conditional inference for mixed-effects models to GAMs based on the penalized log-likelihood (\ref{EQ:penalized_log_likelihood_GAM_LAML_November_28_2022}). The resulting confidence intervals (CIs) of $\Psi$'s depend on the conditional MSEs of $\theta$ estimates, which are equal to the corresponding marginal MSEs, namely, the variance of MMLEs, as referred above. 

The objective of this work is to develop interval estimates for REs in the conditional setting that have accurate coverage probabilities. We refer to these RE interval estimates as conditional CIs instead of prediction intervals (PIs), which is the conventional terminology for REs, because the REs are actually fixed in the conditional inference setting. In Sec. \ref{sec:Method}, we will first identify the reason why the commonly used conditional CIs based on the MSEs of the posterior mode RE estimates do not perform well, and then propose improved interval estimators accordingly. In Sec. \ref{sec:simulation}, we will use simulation studies to examine the performance of the new interval estimators for mixed-effects models and GAMs, and compare them with the interval estimators based on marginal variances that are typically used. Discussions and concluding remarks are provided in Sec. \ref{sec:discussion}.

\section{Methods}\label{sec:Method}
\subsection{Notations and background}\label{sec:Notations_background}
Let $\Psi$ denote the vector of REs in the nonlinear mixed effects model described by (\ref{EQ:lj_November_28_2022}). $\Psi$ can also represent the basis coefficients in a GAM, because of the equivalence between mixed-effects models in the conditional setting and GAMs, as reflected by (\ref{EQ:lj_November_28_2022}) and (\ref{EQ:penalized_log_likelihood_GAM_LAML_November_28_2022}).
The model parameters are denoted as $\theta$.
For GAMs, $\theta$ also includes the smoothing parameters.
$\theta$ are estimated by maximizing the marginal likelihood, which is the marginal distribution of data $D$ obtained by integrating out REs $\Psi$ from the joint distribution of $\Psi$ and $D$. This MMLE is denoted as $\hat{\theta}$. $\Psi$ are estimated by maximizing the log-joint likelihood (\ref{EQ:lj_November_28_2022}), or equivalently the log-posterior distribution of $\Psi$, with $\hat{\theta}$ used for $\theta$. This log-joint likelihood formally includes the GAM penalized log-likelihood (\ref{EQ:penalized_log_likelihood_GAM_LAML_November_28_2022}). The posterior mode estimators of $\Psi$ are denoted as $\hat{\Psi}(\hat{\theta})$.

If the distribution of $\Psi$ conditional on observed data is approximately multivariate normal (MVN), the conditional bias and covariance of $\hat{\Psi}(\hat{\theta})$ are given by \citep{Zheng2021P2}
\begin{linenomath*} 
	\begin{align}\label{EQ:bias_hatpsi_f_December_6_2022}
		\begin{split}
			\mathrm{E}\{ \hat{\Psi}(\hat{\theta})\,|\,\Psi\} - \Psi &=
			-\ddot{l}_{j}(\Psi,\theta_o)^{-1} \ddot{l}_r(\Psi,\theta_o) [\Psi-\mathrm{E}\{\Psi\}] + O(T^{-1/2}),\\
			\mathrm{Cov}\{ \hat{\Psi}(\hat{\theta})\,|\,\Psi\}  &\approx
			-\ddot{l}_{j}(\Psi,\theta_o)^{-1} + \ddot{l}_{j}(\Psi,\theta_o)^{-1}\ddot{l}_r(\Psi,\theta_o) \ddot{l}_{j}(\Psi,\theta_o)^{-1}\\
			&\hspace{5mm}+ \dfrac{\partial\hat{\Psi}(\theta_o)}{\partial\theta_o^{\top}}\mathrm{Cov}(\hat{\theta}\,|\,\Psi) \dfrac{\partial\hat{\Psi}^{\top}(\theta_o)}{\partial\theta_o}, 
		\end{split}	
	\end{align} 
\end{linenomath*}\begin{comment}
	Here the conditional covariance of $\hat{\theta}$ is given by \citep{Zheng2021P2}
	\begin{linenomath*} 
		\begin{align}\label{cov_theta_hat_conditional_on_Psi_March_10_2021}
			\begin{split}
				\mathrm{Cov}(\hat{\theta}\,|\,\Psi) &=-\ddot{l}_m^{-1}(\theta_o) - \ddot{l}_m^{-1}(\theta_o) \, \widetilde{\mathcal{I}}_r\, \ddot{l}_m^{-1}(\theta_o) + o(T^{-1}),
			\end{split}	
		\end{align}
	\end{linenomath*} 
	where $l_m$ is the marginal likelihood obtained by integrating the joint likelihood $l_j$ across REs $\Psi$, $\ddot{l}_m(\theta)$ denotes its second order derivative respecting parameter $\theta$, and 
	\begin{linenomath*} 
		\begin{align*}
			\begin{split}
				\widetilde{\mathcal{I}}_r &= -\dfrac{\partial^2 l_r(\Psi,\theta_o)}{\partial\theta_o\partial\theta_o^{\top}} - \dfrac{\partial^2 l_r(\Psi,\theta_o)}{\partial\theta_o\partial\Psi^{\top}}\dfrac{\partial\hat{\Psi}(\hat{\theta})}{\partial\hat{\theta}^{\top}} - \dfrac{\partial\hat{\Psi}^{\top}(\hat{\theta})}{\partial\hat{\theta}}\dfrac{\partial^2 l_r(\Psi,\theta_o)}{\partial\Psi\partial\theta_o^{\top}}\\
				&\hspace{4.5mm} - \dfrac{\partial\hat{\Psi}^{\top}(\hat{\theta})}{\partial\hat{\theta}}\dfrac{\partial^2 l_r(\Psi,\theta_o)}{\partial\Psi\partial\Psi^{\top}}\dfrac{\partial\hat{\Psi}(\hat{\theta})}{\partial\hat{\theta}^{\top}}.
			\end{split}			
		\end{align*}	
\end{linenomath*} \end{comment}
where the double dots above $\l_j$ and $l_r$ denote second order derivatives with respect to $\Psi$. $\mathrm{Cov}(\hat{\theta}\,|\,\Psi)$ is of $O(T^{-1})$ \citep{Zheng2021P2}, but its explicit form is not relevant to this paper. \begin{comment}Because the estimates $\hat{\theta}$ and $\hat{\Psi}$ are used in place of their true values, the estimator $\ddot{l}_m^{-1}(\hat{\theta}) \, \widetilde{\mathcal{I}}_r(\hat{\theta},\hat{\Psi})\, \ddot{l}_m^{-1}(\hat{\theta})$ for the second term in (\ref{cov_theta_hat_conditional_on_Psi_March_10_2021}) may not be positive definite, wherein \cite{Zheng2021P2} recommend to use its nearest positive definite matrix \citep{higham2002computing}. \end{comment}

When deriving the conditional MSE of $\hat{\Psi}(\hat{\theta})$ given $\Psi$, the bias-squared term $\ddot{l}_{j}(\Psi,\theta_o)^{-1} \ddot{l}_r(\Psi,\theta_o) [\Psi-\mathrm{E}\{\Psi\}][\Psi^{\top}-\mathrm{E}\{\Psi\}^{\top}] \ddot{l}_r(\Psi,\theta_o)\ddot{l}_{j}(\Psi,\theta_o)^{-1}$ cannot be evaluated correctly because $\Psi$ is unknown and can only be estimated with bias of $O(1)$ according to (\ref{EQ:bias_hatpsi_f_December_6_2022}).
Let $\Omega=(\Psi^{\top},\theta_o^{\top})^{\top}$ and $\hat{\Omega}=(\hat{\Psi}(\hat{\theta})^{\top},\hat{\theta}^{\top})^{\top}$. \citet[Appendix D]{Zheng2021P2} showed that if the bias-squared term is approximated by its marginal expectation, then the conditional MSE is approximately
\begin{linenomath*} 
	\begin{align}\label{EQ:mse_hatpsi_f_approx}
		\begin{split}
			&\mathrm{MSE}( \hat{\Omega}\,|\,\Psi) = \mathrm{Cov}( \hat{\Omega}\,|\,\Psi) + \mathrm{E}( \hat{\Omega}-\Omega\,|\,\Psi) \mathrm{E}( \hat{\Omega}-\Omega\,|\,\Psi)^{\top}\\
			&\approx \mathrm{Cov}( \hat{\Omega}\,|\,\Psi) + \mathrm{E}\left\lbrace \mathrm{E}(\hat{\Omega}-\Omega\,|\,\Psi)\, \mathrm{E}(\hat{\Omega}-\Omega\,|\,\Psi)^{\top} \right\rbrace  \\
			&= \left[ \begin{array}{cc}
				-\ddot{l}_j^{-1} + \dfrac{\partial\hat{\Psi}(\hat{\theta})}{\partial\hat{\theta}^{\top}} \mathrm{Cov}( \hat{\theta} ) \dfrac{\partial\hat{\Psi}^{\top}(\hat{\theta})}{\partial\hat{\theta}} & \dfrac{\partial\hat{\Psi}(\hat{\theta})}{\partial\hat{\theta}^{\top}} \mathrm{Cov}( \hat{\theta} ) \\
				\mathrm{Cov}( \hat{\theta} ) \dfrac{\partial\hat{\Psi}^{\top}(\hat{\theta})}{\partial\hat{\theta}}	& \mathrm{Cov}( \hat{\theta} )
			\end{array}
			\right]+o(1/T).
		\end{split}	
	\end{align}
\end{linenomath*} 
The last matrix in (\ref{EQ:mse_hatpsi_f_approx}) is the MSE in the marginal setting \citep{zheng2021frequentist} and is also the posterior variability in empirical Bayes \citep{kass1989approximate}.
Note that in this conditional setting, $\hat{\theta}$ has a bias of order $O(T^{-1/2})$.
The conditional MSE of $\hat{\theta}$ is approximately the marginal covariance $\mathrm{Cov}( \hat{\theta} )=-\ddot{l}_m^{-1}(\hat{\theta})$, where $l_m$ is the logarithm of the marginal likelihood obtained by integrating the joint likelihood across REs $\Psi$, and $\ddot{l}_m(\theta)$ denotes its second order derivatives with respect to $\theta$.  

The approximation made in (\ref{EQ:mse_hatpsi_f_approx}) indicates that when applying the marginal MSE (or equivalently the Bayesian posterior variance) to construct CIs in the conditional setting, we actually approximate the bias-squared term by integrating it across the marginal distribution of REs, which introduces biases of order $O(1)$ to the conditional MSE. As a result, the corresponding component-wise CI coverage probabilities can be unsatisfactory. When the CI coverage probabilities are averaged over all the REs in the model, conditional MSEs, and hence bias-squared terms, are somewhat averaged over REs approximately, which in turn is similar to an average over the marginal distribution of REs based on the notion of ergodicity, i.e., the ensemble average equals the time average \citep{feller2008introduction}. Therefore, the coverage probability averaged across the REs can be substantially closer to the nominal level than the component-wise CI coverage probabilities, as discussed for smoothing splines in \cite{nychka1988bayesian}.

\subsection{Conditional CI based on bias-corrected RE estimation}\label{sec:conditional_CI_December_19_2022}
The bias-squared term in the conditional MSE cannot be estimated well and hence adversely affects the component-wise CI coverage.
Using a bias corrected RE estimator can avoid this term in the conditional MSE and hence improve CI coverage probabilities.
To simplify notation we write the bias formula in (\ref{EQ:bias_hatpsi_f_December_6_2022}) as
\begin{linenomath*} 
	\begin{align}\label{EQ:bias_hatpsi_f_December_20_2022}
		\begin{split}
			\mathrm{E}\{ \hat{\Psi}(\hat{\theta})\,|\,\Psi\} - \ddot{l}_{j}(\Psi,\theta_o)^{-1} \ddot{l}_r(\Psi,\theta_o) \mathrm{E}\{ \Psi \} &= \mathbf{B}\,\Psi + O(T^{-1/2}), 
		\end{split}	
	\end{align} 
\end{linenomath*}
where $\mathbf{B}=\pmb{\mathrm{I}}
-\ddot{l}_{j}(\Psi,\theta_o)^{-1} \ddot{l}_r(\Psi,\theta_o)$ and $\pmb{\mathrm{I}}$ is an identity matrix. The $O(T^{-1/2})$ term comes from the expectation of $\partial\hat{\Psi}(\theta_o)/\partial\theta_o^{\top}(\hat{\theta}-\theta_o)$. If some components of $\partial\hat{\Psi}(\theta_o)/\partial\theta_o^{\top}(\hat{\theta}-\theta_o)$ are large, i.e., close to 1 in absolute value, this term can become large. Therefore, we make the assumption in our subsequent development that $\partial\hat{\Psi}(\theta_o)/\partial\theta_o^{\top}$ is not large. Reparameterization can often resolve the issue of large components of $\partial\hat{\Psi}(\theta_o)/\partial\theta_o^{\top}$ in cases where they occur. For instance, if $\mu + \Psi$ follows an AR(1) model with mean $\mu$, then the components of $\partial\hat{\Psi}/\partial\mu$ are close to 1. However, if we redefine $\Psi$ to follow an AR(1) model with mean $\mu$, then the components of $\partial\hat{\Psi}/\partial\mu$ become small. 

%For simplicity, we only consider the case when $\mathrm{E}\{\Psi\}=0$, which is quite common when implementing mixed-effects models. The following derivations are straight-forward to extend if $\mathrm{E}\{\Psi\}\neq0$. 

If all the REs are well supported by data and hence $\mathbf{B}$ is nonsingular, multiplying both sides of (\ref{EQ:bias_hatpsi_f_December_20_2022}) by $\mathbf{B}^{-1}$ does not change the approximation order $O(T^{-1/2})$, and
\begin{linenomath*} 
	\begin{align}\label{EQ:bias_corrected_December_20_2022}
		\begin{split}
			\hat{\Psi}_{\mathrm{BC}}(\hat{\theta}) = \mathbf{B}^{-1} \left[ \hat{\Psi}(\hat{\theta}) - \ddot{l}_{j}(\Psi,\theta_o)^{-1} \ddot{l}_r(\Psi,\theta_o)\, \mathrm{E}\{ \Psi \} \right] 
		\end{split}	
	\end{align} 
\end{linenomath*}
is an unbiased estimator of $\Psi$ to the order $O(T^{-1/2})$. Here the subscript ``BC" denotes ``bias correction", and $\mathrm{E}\{ \Psi \}$ as a function of $\theta$ is evaluated at $\hat{\theta}$. In (\ref{EQ:bias_corrected_December_20_2022}) $\ddot{l}_{j}(\Psi,\theta_o)$ and $\ddot{l}_r(\Psi,\theta_o)$ are all evaluated at $\hat{\Psi}(\hat{\theta})$ and $\hat{\theta}$, which does not change the approximation order $O(T^{-1/2})$ because $\hat{\theta}-\theta_o$ is $O_p(T^{-1/2})$ as the MSE of $\hat{\theta}$ is $O(T^{-1})$, and $l_{j}$ and $l_r$ are approximately MVN respecting $\Psi$. In Sec. \ref{sec:conditional_CI_no_data_December_20_2022} we extend (\ref{EQ:bias_corrected_December_20_2022}) when some REs are not well supported by the data so that $\mathbf{B}$ is singular or close to singular. 

If $\hat{\Psi}_{\mathrm{BC}}(\hat{\theta})$ is an unbiased estimator of $\Psi$ then its conditional MSE is equal to its conditional covariance,
\begin{linenomath*} 
	\begin{align}\label{EQ:mse_hat_Psi_BC_December_22_2022}
		\begin{split}
			&\mathrm{MSE}( \hat{\Psi}_{\mathrm{BC}}(\hat{\theta})\,|\,\Psi) = \mathrm{Cov}( \hat{\Psi}_{\mathrm{BC}}(\hat{\theta})\,|\,\Psi) + O(T^{-1})\\
			&= \mathbf{B}^{-1}\left[ -\ddot{l}_{j}(\Psi,\theta_o)^{-1} + \ddot{l}_{j}(\Psi,\theta_o)^{-1}\ddot{l}_r(\Psi,\theta_o) \ddot{l}_{j}(\Psi,\theta_o)^{-1} \right. \\
			& \hspace{15mm} \left. + \Upsilon\,\mathrm{Cov}(\hat{\theta})\, \Upsilon^{\top} \right] \left\lbrace \mathbf{B}^{-1} \right\rbrace^{\top}+ O(T^{-1}),\\
			&\Upsilon = \dfrac{\partial\hat{\Psi}(\theta_o)}{\partial\theta_o^{\top}} - \ddot{l}_{j}(\Psi,\theta_o)^{-1} \ddot{l}_r(\Psi,\theta_o) \dfrac{\partial\mathrm{E}\{ \Psi \}}{\partial\theta_o^{\top}} .  
		\end{split}	
	\end{align}
\end{linenomath*} 
Here $\mathrm{Cov}(\hat{\theta}\,|\,\Psi)$ in (\ref{EQ:bias_hatpsi_f_December_6_2022}) is replaced by $\mathrm{Cov}(\hat{\theta})$ because conditional on $\Psi$ $\hat{\theta}$ is a biased estimator of $\theta_o$ and hence its overall variability should be measured by its MSE $\mathrm{Cov}(\hat{\theta})$ (see Eq. \ref{EQ:mse_hatpsi_f_approx}). When 
applying other estimators of $\theta$ (e.g., Eq. \ref{EQ:penalized_log_likelihood_GAM_November_28_2022}, the penalized likelihood maximization in GAM), $\mathrm{Cov}(\hat{\theta})$ in (\ref{EQ:mse_hat_Psi_BC_December_22_2022}) should be replaced by the corresponding MSEs.
%where $\mathbf{A} = \mathbf{B}^{-1} [ \partial\hat{\Psi}(\theta_o)/\partial\theta_o^{\top} - \ddot{l}_{j}(\Psi,\theta_o)^{-1} \ddot{l}_r(\Psi,\theta_o)\, \partial\mathrm{E}\{\Psi\}/\partial\theta_o^{\top} ]$. 
A $100(1-\alpha)\%$ CI for the $i$th component of $\Psi$, $\Psi_i$, can be constructed as
\begin{linenomath*} 
	\begin{align}\label{EQ:CI_Psi_October_23_2022}
		\hat{\Psi}_{\mathrm{BC},i}(\hat{\theta}) \pm z_{\alpha/2}\,\hat{\sigma}_{\Psi,i},
	\end{align}
\end{linenomath*}
where $\hat{\sigma}_{\Psi,i}=\sqrt{\mathrm{MSE}( \hat{\Psi}_{\mathrm{BC}}(\hat{\theta})\,|\,\Psi)_{ii}}$, $z_{\alpha/2}$ is the $z$-value corresponding to an area $\alpha/2$ in the upper tail of a standard normal distribution, and the subscript $i$ denotes the $i$th element. 

A CI for a vector-valued function of $\Psi$, $\mathbf{g}(\Psi)$, can be constructed with the generalized-delta method \cite[see, e.g.,][]{kristensen2015tmb}, where $\mathbf{g}(\Psi)$ is regarded as a MVN random variable with mean $\hat{\mu}_{\mathbf{g}} = \mathbf{g}(\hat{\Psi}_{\mathrm{BC}}(\hat{\theta}))$ and covariance $\hat{\Sigma}_{\mathbf{g}}=\{\partial\mathbf{g}(\Psi)/\partial\Psi^{\top}\}\mathrm{MSE}( \hat{\Psi}_{\mathrm{BC}}(\hat{\theta})\,|\,\Psi)\{\partial\mathbf{g}^{\top}(\Psi)/\partial\Psi\}$.
When $\mathbf{g}(\cdot)$ is a nonlinear function, a better estimator of its mean $\hat{\mu}_{\mathbf{g}}$ can be obtained through the bias correction proposed in \cite{thorson2016implementing}. A $100(1-\alpha)\%$ CI for the $i$th component of $\mathbf{g}(\Psi)$ is
\begin{linenomath*} 
	\begin{align}\label{EQ:CI_g_October_23_2022}
		\hat{\mu}_{\mathbf{g},i}\pm z_{\alpha/2}\,\hat{\sigma}_{\mathbf{g},i},
	\end{align}
\end{linenomath*}
where $\hat{\sigma}_{\mathbf{g},i}=\sqrt{\hat{\Sigma}_{\mathbf{g},ii}}$. The generalized-delta method is similarly implemented with the other $\Psi$ estimators proposed in this paper.

\subsection{Conditional CI when some REs are not associated with data}\label{sec:conditional_CI_no_data_December_20_2022}
Let $\pmb{0}$ denote a matrix or vector of zeros with the appropriate dimension. If there is no data, namely $\ddot{l}_c(\Psi,\theta_o)= \ddot{l}_j(\Psi,\theta_o) - \ddot{l}_r(\Psi,\theta_o) =\pmb{0}$, $\ddot{l}_{j}(\Psi,\theta_o)^{-1} \ddot{l}_r(\Psi,\theta_o)=\pmb{\mathrm{I}}$ in (\ref{EQ:bias_hatpsi_f_December_20_2022}) and hence $\mathbf{B}=\pmb{0}$.
In this case, $\hat{\Psi}(\hat{\theta})$ is zero.
When we have no information about REs then we have to use the marginal MSE to provide CIs if we have enough knowledge about the model parameters.
Even if only a small number of REs are not supported by data, which will commonly occur if there are missing data or other types of irregular sampling, then $\mathbf{B}$ will be singular.

We need to invert $\mathbf{B}$ in (\ref{EQ:bias_hatpsi_f_December_20_2022}) for our unbiased estimator of the REs (i.e., Eq. \ref{EQ:bias_corrected_December_20_2022}).
However, when REs associated with few or no data cause singularity in $\mathbf{B}$, it is not possible to implement (\ref{EQ:bias_corrected_December_20_2022}) directly. In this case, a straightforward remedy for those REs not supported by data is to use the corresponding components of $\hat{\Psi}(\hat{\theta})$ for $\hat{\Psi}_{\mathrm{BC}}(\hat{\theta})$, with uncertainty given by the marginal MSE (\ref{EQ:mse_hatpsi_f_approx}).
We consider this further in the Discussion Section.
Therefore, we first devise a method to identify whether the REs are supported by data sufficiently or not, and then for those REs with sufficient data we use (\ref{EQ:bias_corrected_December_20_2022}) and (\ref{EQ:mse_hat_Psi_BC_December_22_2022}) to construct CIs, and for those REs with insufficient data we use $\hat{\Psi}(\hat{\theta})$ and (\ref{EQ:mse_hatpsi_f_approx}).

We use the singular value decomposition, $\mathbf{B}=U\Gamma V^{\top}$, where $U$ and $V$ are orthogonal matrices, and $\Gamma$ is a diagonal matrix of non-negative singular values. Multiplying both sides of (\ref{EQ:bias_hatpsi_f_December_20_2022}) with $U^{\top}$ gives
\begin{linenomath*} 
	\begin{align*}
		\mathrm{E}\{U^{\top}[ \hat{\Psi}(\hat{\theta}) - \ddot{l}_{j}(\Psi,\theta_o)^{-1} \ddot{l}_r(\Psi,\theta_o)\, \mathrm{E}\{ \Psi \} ]\,|\,\Psi\}&=\Gamma\, V^{\top}\Psi + O(T^{-1/2}).
	\end{align*}
\end{linenomath*}
Here we applied $U^{\top}O(T^{-1/2}) = O(T^{-1/2})$.
$V^{\top}\Psi$ is an alternative set of REs to the original $\Psi$, and is estimated by $U^{\top}[ \hat{\Psi}(\hat{\theta}) - \ddot{l}_{j}(\Psi,\theta_o)^{-1} \ddot{l}_r(\Psi,\theta_o)\, \mathrm{E}\{ \Psi \} ]$. Component-wise, $\mathrm{E}\{[U^{\top}[ \hat{\Psi}(\hat{\theta}) - \ddot{l}_{j}(\Psi,\theta_o)^{-1} \ddot{l}_r(\Psi,\theta_o)\, \mathrm{E}\{ \Psi \} ]]_i|\Psi\} =\Gamma_{ii}\, [V^{\top}\Psi]_i + O(T^{-1/2})$, where the subscript $i$ denotes the $i$th element. If multiplying both sides with $\Gamma_{ii}^{-1}$ does not change the approximation order $O(T^{-1/2})$, namely, if $\Gamma_{ii}>\gamma_c$, say, then $\Gamma_{ii}^{-1}[U^{\top}[ \hat{\Psi}(\hat{\theta}) - \ddot{l}_{j}(\Psi,\theta_o)^{-1} \ddot{l}_r(\Psi,\theta_o)\, \mathrm{E}\{ \Psi \} ]]_i$ is approximately an unbiased estimator of $[V^{\top}\Psi]_i$; otherwise, $[V^{\top}\Psi]_i$ is estimated by $[V^{\top}\hat{\Psi}(\hat{\theta})]_i$ with a bias $-[V^{\top}\ddot{l}^{-1}_j\ddot{l}_r(\Psi-\mathrm{E}\{ \Psi \})]_i$, according to (\ref{EQ:bias_hatpsi_f_December_6_2022}). This component-by-component treatment leads to an estimator of $V^{\top}\Psi$, $[\Gamma_g^{-1}U^{\top} + \Gamma^c\, V^{\top}] \hat{\Psi}(\hat{\theta}) - \Gamma_g^{-1}U^{\top} \ddot{l}_{j}(\Psi,\theta_o)^{-1} \ddot{l}_r(\Psi,\theta_o)\, \mathrm{E}\{ \Psi \} $, with a bias $-\Gamma^c\, V^{\top}\ddot{l}^{-1}_j\ddot{l}_r(\Psi-\mathrm{E}\{ \Psi \})$, where $\Gamma^c$ is a diagonal matrix with value 1 replacing the zero or very small singular values ($\leq\gamma_c$) in $\Gamma$, but 0 otherwise, and $\Gamma_g^{-1}$ is similar to the generalized inverse \citep{ben2003generalized} of $\Gamma$, namely, setting very small singular values ($\leq\gamma_c$) to zero, but inverting the other diagonal values. This estimator of $V^{\top}\Psi$ is then transformed to an estimator of $\Psi$, $\hat{\Psi}_{\mathrm{SD}}(\hat{\theta}) =V[(\Gamma_g^{-1}U^{\top} + \Gamma^c\, V^{\top}) \hat{\Psi}(\hat{\theta}) - \Gamma_g^{-1}U^{\top}\ddot{l}_{j}(\Psi,\theta_o)^{-1} \ddot{l}_r(\Psi,\theta_o)\, \mathrm{E}\{ \Psi \} ]$, with
\begin{linenomath*} \label{EQ:Psi_BC_bias}
	\begin{align}
		\mathrm{E}\{\hat{\Psi}_{\mathrm{SD}}(\hat{\theta}) |\Psi\}-\Psi&=-V\Gamma^c\, V^{\top}\ddot{l}^{-1}_j\ddot{l}_r\left( \Psi-\mathrm{E}\{ \Psi \}\right)  + O(T^{-1/2}).
	\end{align}
\end{linenomath*}
Here the subscript ``SD" denotes ``singular value decomposition".

The MSE of $\hat{\Psi}_{\mathrm{SD}}(\hat{\theta})$ is 
\begin{linenomath*} 
	\begin{align}\label{EQ:Psi_BC_mse}
		\begin{split}
			\mathrm{MSE}(\hat{\Psi}_{\mathrm{SD}}(\hat{\theta})|\Psi)&=V\left[ \Gamma_g^{-1}U^{\top}  \mathrm{Cov}_{11}\,U\Gamma_g^{-1} + \Gamma_g^{-1}U^{\top}  \mathrm{Cov}_{12}\,V\Gamma^c\right. \\ &\hspace{10mm}\left. + \Gamma^c\, V^{\top}  \mathrm{Cov}_{12}^{\top}\,U\Gamma_g^{-1} + \Gamma^c\, V^{\top}\mathrm{MSE}(\hat{\Psi}(\hat{\theta})|\Psi)\,V\Gamma^c\right] V^{\top},			
		\end{split}		
	\end{align}
\end{linenomath*}
where using $\Upsilon$ defined in (\ref{EQ:mse_hat_Psi_BC_December_22_2022}),
\begin{linenomath*} 
	\begin{align}\label{EQ:cov_hat_Psi_conditional_April_11_2023}
		\begin{split}
			\mathrm{Cov}_{11} &= -\ddot{l}_{j}(\Psi,\theta_o)^{-1} + \ddot{l}_{j}(\Psi,\theta_o)^{-1}\ddot{l}_r(\Psi,\theta_o) \ddot{l}_{j}(\Psi,\theta_o)^{-1} + \Upsilon\,\mathrm{Cov}(\hat{\theta})\, \Upsilon^{\top},\\
			\mathrm{Cov}_{12} &= -\ddot{l}_{j}(\Psi,\theta_o)^{-1} + \ddot{l}_{j}(\Psi,\theta_o)^{-1}\ddot{l}_r(\Psi,\theta_o) \ddot{l}_{j}(\Psi,\theta_o)^{-1} + \Upsilon\,\mathrm{Cov}(\hat{\theta}) \dfrac{\partial\hat{\Psi}^{\top}(\theta_o)}{\partial\theta_o}.   
		\end{split}	
	\end{align}
\end{linenomath*} 
Eq. (\ref{EQ:Psi_BC_mse}) says that the conditional MSE for those elements of $\Psi$ with few data is the marginal MSE, and for the other REs we use the conditional MSE for bias-corrected estimators (\ref{EQ:mse_hat_Psi_BC_December_22_2022}). The $100(1-\alpha)\%$ CI for $\Psi_i$ can be constructed as $\hat{\Psi}_{\mathrm{SD},i}(\hat{\theta}) \pm z_{\alpha/2}\sqrt{\mathrm{MSE}( \hat{\Psi}_{\mathrm{SD}}(\hat{\theta})\,|\,\Psi)_{ii}}$. When $\gamma_c$ goes to zero, this CI converges to the conditional CI based on the bias-corrected estimator (\ref{EQ:CI_Psi_October_23_2022}). When $\gamma_c$ becomes large, this CI converges to the CI based on the marginal MSE. The REs are usually defined for observational units \citep{kass1989approximate, flores2019bootstrap}. To determine $\gamma_c$, we can count the number $n_c$ of observational units that have data, and $\gamma_c$ is specified so that exactly $n_c$ singular values exceed $\gamma_c$. If different types of data are involved, e.g., survey catches and fisheries catches in a fisheries stock assessment model, then $n_c$ should be the summation of the observational units with data for the various types. $n_c$ is also equal to the number of nonzero singular values from the singular value decomposition of $\ddot{l}_c$.

We use $\hat{\Omega}_{\mathrm{SD}}=(\hat{\Psi}_{\mathrm{SD}}(\hat{\theta})^{\top},\hat{\theta}^{\top})^{\top}$ to estimate $\Omega=(\Psi^{\top},\theta_o^{\top})^{\top}$ for the case with data insufficiency. The conditional MSE of $\hat{\Omega}_{\mathrm{SD}}$ is given by
\begin{linenomath*} 
	\begin{align}\label{EQ:mse_SVD_hatpsi_f_approx}
		\begin{split}
			&\mathrm{MSE}( \hat{\Omega}_{\mathrm{SD}}\,|\,\Psi) = \mathrm{MSE}\left(  \left[ \begin{array}{c}
				\hat{\Psi}_{\mathrm{SD}}(\hat{\theta}) \\
				\hat{\theta}
			\end{array}
			\right] \,\middle|\,\Psi\right) \\			
			&= \left[ \begin{array}{cc}
				\mathrm{MSE}(\hat{\Psi}_{\mathrm{SD}}(\hat{\theta})|\Psi) & G\,\mathrm{Cov}( \hat{\theta} ) \\
				\mathrm{Cov}( \hat{\theta} ) \,G^{\top}	& \mathrm{Cov}( \hat{\theta} )
			\end{array}
			\right],
		\end{split}	
	\end{align}
\end{linenomath*} 
where $G=V[(\Gamma_g^{-1}U^{\top} + \Gamma^c\, V^{\top})\, \partial\hat{\Psi}(\theta_o)/\partial\theta_o^{\top} - \Gamma_g^{-1}U^{\top}\ddot{l}_{j}(\Psi,\theta_o)^{-1} \ddot{l}_r(\Psi,\theta_o)\, \partial\mathrm{E}\{ \Psi \}/\partial\theta_o^{\top} ]$.

\section{Simulation studies}\label{sec:simulation} 
In this section we use a random walk example and one based on the global temperature anomaly data in \citet[Fig. 3]{wood2020inference} to examine the performance of the conditional CIs derived from the bias-corrected RE estimation in Sec. \ref{sec:conditional_CI_December_19_2022}. We also examine the impact of missing data on CI performance.

\subsection{Random walk example}\label{sec:random_walk_December_28_2022} 

The random walk is $\Psi_t|\Psi_{t-1} \stackrel{indep}{\sim} N(\Psi_{t-1},\sigma^{2}_{\Psi})$ for $t= 2,...,T$, and $\Psi_{1} \stackrel{indep}{\sim} N(0,\sigma^{2}_{\Psi})$. 
Here $N(\mu,\sigma^2)$ denotes the normal distribution with mean $\mu$ and variance $\sigma^2$.
At each time-step there are $n$ independent observations of the process, $Y_{t,i}|\Psi_t \stackrel{indep}{\sim} N(\Psi_t,\sigma^{2}_{\epsilon}),\, i=1,...,n$ and $t=1,...,T$.
The parameters are $\theta = (\sigma_{\Psi},\sigma_{\epsilon})^{\top}$ and the REs are $\Psi = (\Psi_{1},...,\Psi_{T})^{\top}$ which is a $T \times 1$ vector.
We randomly generated one set of random-walk $\Psi_t$'s, and then generate 1000 simulations of the data $y$ conditional on the $\Psi_t$'s, with $\sigma_{\Psi} = 1$, $\sigma_{\epsilon} = 0.5$, and two choices each for $n=2,5$ and $T=50,200$.
We estimated the parameters and $\Psi_t$'s using TMB \citep{kristensen2015tmb} and the nlminb procedure in R \citep{r2018citation}.
Biases and CI coverage depend on the values of $\Psi_t$'s, and to generalize results we also repeated the simulations for 500 different random walk $\Psi$'s, and then averaged results over these random $\Psi$'s.

\begin{figure}[h]
	\centering
	\includegraphics{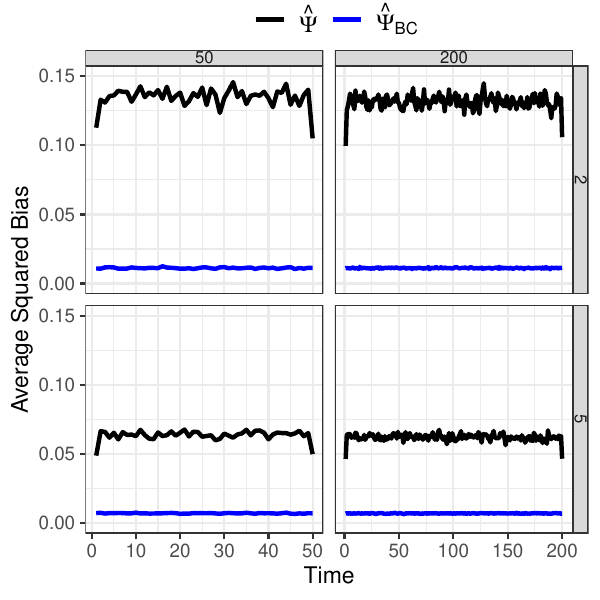}
	%\vspace*{-5mm}
	\caption{Simulated squared bias for estimators of the random walk $\Psi$'s, averaged over 500 simulated $\Psi$'s. For each set of $\Psi$'s, the squared bias is based on 1000 simulated data sets $Y_{t,i}|\Psi$ with $i=1,...,n$ and $t = 1,...,T$. Panel rows indicate choices for $n$ and columns indicate $T$. The colors correspond to the ``posterior mode" estimator ($\hat{\Psi}$) and the bias-corrected estimator ($\hat{\Psi}_{\mathrm{BC}}$).}
	\label{fig:Bias2_RW_EX}	
\end{figure}

\begin{figure}[h]
	\centering
	\includegraphics{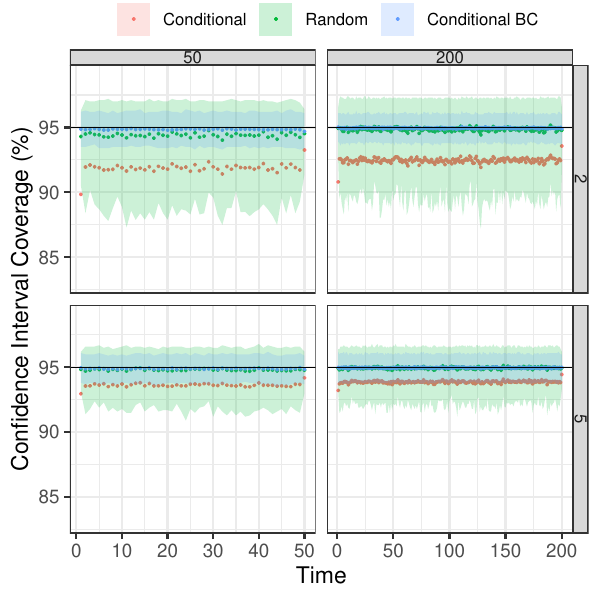}
	%	\vspace*{-5mm}
	\caption{Simulated coverage of 95\% CIs for the random walk $\Psi$'s.
		Points indicate average CI coverages, and shaded regions indicate 5\% and 95\% quantiles, from the 500 simulated $\Psi$'s.
		Panels are described in Fig. \ref{fig:Bias2_RW_EX}.}
	\label{fig:CI_RW}	
\end{figure}

The squared bias, $(\mathrm{E}\{\hat{\Psi}_{BC}|\Psi\} - \Psi)^2$, was much smaller than for $\hat{\Psi}$, when averaged over the 500 simulation $\Psi$'s (Fig. \ref{fig:Bias2_RW_EX}).
This indicates that the bias correction is effective.
In Fig. \ref{fig:CI_RW} the ``Random" results are the simulated coverage probabilities of 95\% CIs based on $\hat{\Psi}_i \pm z_{\alpha/2} \times \mathrm{RMMSE}(\hat{\Psi}_i)$, where RMMSE is the root marginal mean squared error (\ref{EQ:mse_hatpsi_f_approx}).
The points are the mean probabilities for the 500 $\Psi$'s, and the green shaded regions indicate the lower $5^{th}$ and upper $95^{th}$ percentiles.
These coverages are accurate on average (with respect to $\Psi$), but for specific values of $\Psi$ the coverages can deviate from 95\%, which is indicated by the width of the shaded regions.
The blue points and shaded regions indicate results for the bias-corrected conditional CI (\ref{EQ:CI_Psi_October_23_2022}).
These coverages are also accurate on average, but more accurate for specific values of $\Psi$ compared to the random intervals.
The red points indicate intervals based on $\hat{\Psi}_{i}$ and the conditional variance (\ref{EQ:bias_hatpsi_f_December_6_2022}).
These are biased because of the conditional bias in $\hat{\Psi}_{i}$.

\begin{figure}[h]
	\centering
	\includegraphics{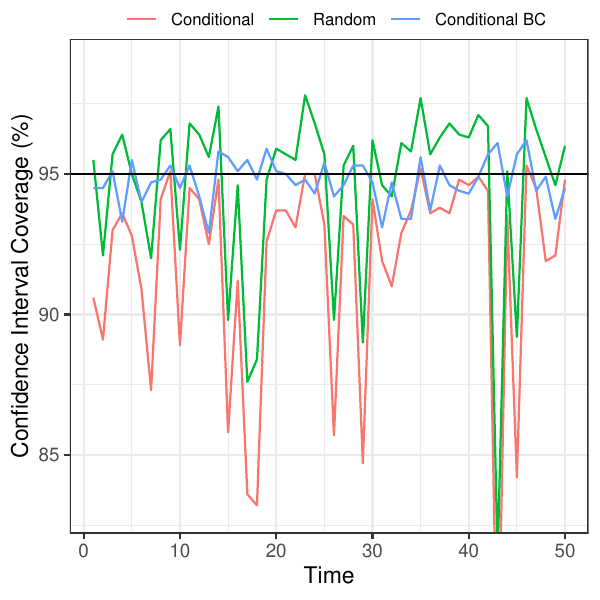}
	%	\vspace*{-5mm}
	\caption{Simulated coverage of 95\% CIs for an illustrative example of random walk $\Psi$'s.
		Red lines indicate CIs based on the "posterior mode" estimator ($\hat{\Psi}$) and conditional-$\Psi$ standard errors (SEs), green lines indicate CIs based on $\hat{\Psi}$ and random-$\Psi$ SEs, and blue lines indicate CIs based on the bias-corrected estimator ($\hat{\Psi}_{\mathrm{BC}}$) and its conditional-$\Psi$ SEs.}
	\label{fig:CI_RW_EX}	
\end{figure}

Simulated coverages for one set of random walk $\Psi$'s are shown in Fig. \ref{fig:CI_RW_EX}.
This figure demonstrates that the conditional bias-corrected CIs have coverages that are close to 95\%, compared to the biased conditional intervals or intervals based on RMMSE.
The latter are the intervals that are commonly used for inferences about random effects in mixed-effects models, and they can be substantially inaccurate and potentially misleading when the random effects are not actually random.

\FloatBarrier
\subsection{Random walk example with missing data}\label{sec:random_walk_missing} 

\begin{figure}[h]
	\centering
	\includegraphics{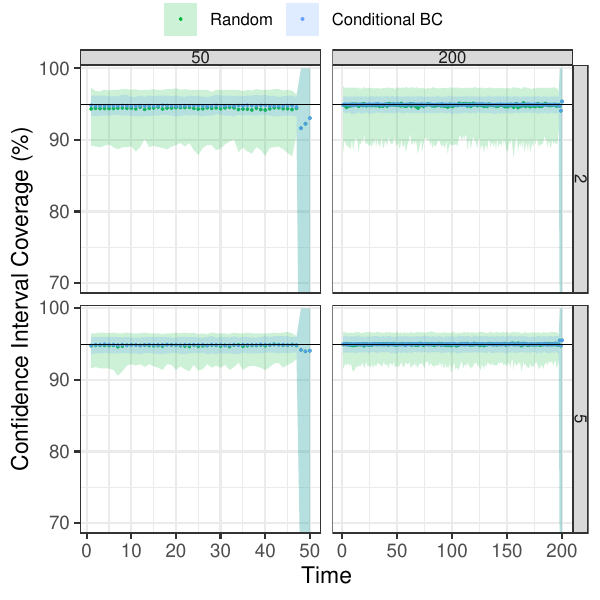}
	%	\vspace*{-5mm}
	\caption{Simulated coverage of 95\% CIs for the random walk $\Psi$'s.
		$y$ observations for the last three time points are missing.
		Points indicate average CI coverages, and shaded regions indicate 5\% and 95\% quantiles, from the 500 simulated $\Psi$'s.
		Panels are described in Fig. \ref{fig:Bias2_RW_EX}.}
	\label{fig:CI_RW_missing}	
\end{figure}

We remove the $y$ observations for the last three random walk time points, and repeated the simulation study described in Section \ref{sec:random_walk_December_28_2022}.
We used the bias-correction procedure outlined in Section \ref{sec:conditional_CI_no_data_December_20_2022} and $\gamma_c=0.1$.
The simulated CI coverages (Fig. \ref{fig:CI_RW_missing}) for time points with data are very similar to those in Fig. \ref{fig:CI_RW}.
Note that the scale of the y-axis is different in these two figures.
However, for the last three time points with no data, the coverages can be substantially different from 95\% for specific $\Psi$'s, and they are only reasonably accurate when averaged over the 500 $\Psi$'s in our simulations.
Note that the random and conditional intervals in Fig. \ref{fig:CI_RW_missing} are identical for the times with no data.

\FloatBarrier
\subsection{Global temperature anomaly example}\label{sec:global_temperature_anomaly_December_28_2022} 
We compare the performances of CIs constructed from $\hat{\Psi}_{\mathrm{BC}}(\hat{\theta})$ and $\hat{\Psi}(\hat{\theta})$ for a GAM applied to the global annual mean temperature anomalies in Fig. 3 of \cite{wood2020inference}. We obtained the global temperature anomalies from \cite{rohde2020berkeley} and used the data during 1850 to 2010, following Fig. 3 of \cite{wood2020inference}. We fit the data using a cubic spline with 50 basis coefficients. The smoothing parameter was estimated by marginal likelihood (REML) maximization.
The annual temperature anomalies were assumed to have independent normal distributions with means given by an intercept plus the cubic spline, and a common standard deviation.
We then assumed that the estimated mean and model parameters were the true values, and used the model to simulate 10000 sets of temperature anomalies. We fit the simulated data with the same model, and constructed CIs using $\hat{\Psi}_{\mathrm{BC}}(\hat{\theta})$ and $\hat{\Psi}(\hat{\theta})$, where the former is based on (\ref{EQ:CI_g_October_23_2022}) and the latter is based on the marginal MSE. The estimates of the smoothing parameter $\lambda$ in (\ref{EQ:penalized_log_likelihood_GAM_LAML_November_28_2022}) were effectively infinite (at a magnitude of $e^{11}$), and hence the usual theory about the variability of MLEs based on the regularity conditions (e.g., the interiority of a parameter to the parameter space) may not apply to the estimation of $\log(\lambda)$ here. When evaluating the variance of the estimator of $\log(\lambda)$ by the inverse of the Hessian of the negative log marginal likelihood with respect to $\log(\lambda)$, too wide CIs based on (\ref{EQ:mse_hat_Psi_BC_December_22_2022}) resulted in our simulation studies. Therefore, we follow the convention of neglecting the uncertainty in $\log(\lambda)$ estimator \citep{wood2016smoothing}. We set up the spline structure using the mgcv package \citep{wood2011fast} in R, and implemented the GAM using TMB.

\begin{figure}[!h]
	\centering
	\includegraphics[width=0.75\linewidth]{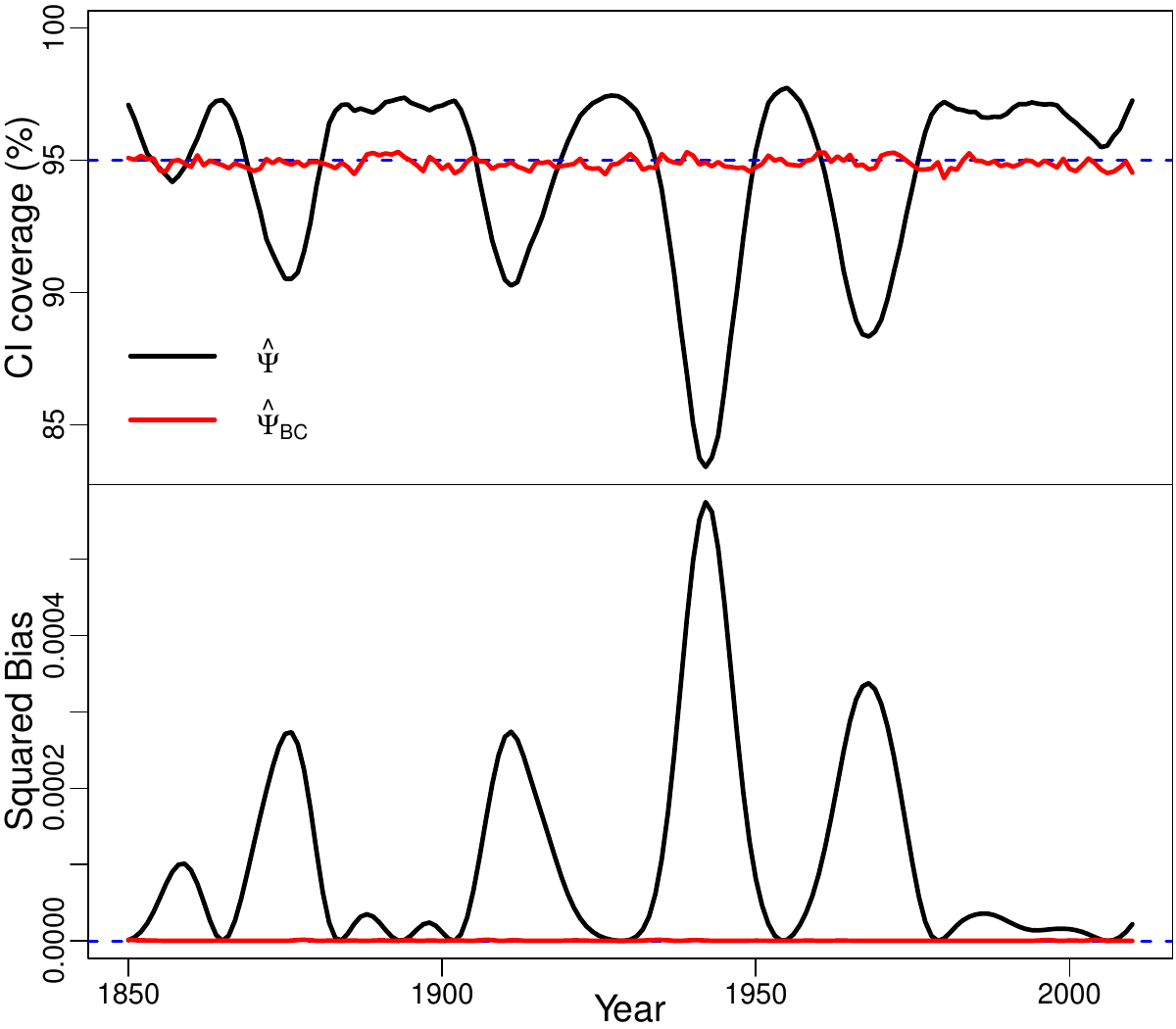}
	\vspace*{-5mm}
	\caption{Simulated 95\% CI coverage rates and squared biases for estimating the yearly mean global temperature anomalies obtained by fitting the global annual temperature anomalies \citep{rohde2020berkeley} using a GAM. Upper panel: the black lines are the coverage rates for the true mean annual temperature anomalies of the 95\% CIs constructed using $\hat{\Psi}(\hat{\theta})$ and marginal MSEs, and the red lines are those based on $\hat{\Psi}_{\mathrm{BC}}(\hat{\theta})$ and (\ref{EQ:CI_g_October_23_2022}); the blue dashed reference line is at 95\%. Lower panel: the black and red lines are the simulated squared biases for estimating the true mean annual temperature anomalies based on $\hat{\Psi}(\hat{\theta})$ and $\hat{\Psi}_{\mathrm{BC}}(\hat{\theta})$ respectively, and the blue dashed reference line is at 0.}
	\label{fig:GAM_annual_temperature_anomalies_January_4_2023}	
\end{figure}

The simulated 95\% CI coverage rates and squared biases are presented in Fig. \ref{fig:GAM_annual_temperature_anomalies_January_4_2023}. The component-wise coverage rates of the CIs based on $\hat{\Psi}_{\mathrm{BC}}(\hat{\theta})$ and (\ref{EQ:CI_g_October_23_2022}) are fairly close to the nominal value, while those based on $\hat{\Psi}(\hat{\theta})$ and marginal MSEs deviate substantially from the nominal value in almost every year.
However, the average CI coverage rate across the years (i.e., across-the-function) for the $\hat{\Psi}(\hat{\theta})$ and marginal MSEs approach is 0.946, close to the nominal value and agreeing with \cite{marra2012coverage} and \cite{wood2020inference}. The across-the-function CI coverage rate for the $\hat{\Psi}_{\mathrm{BC}}(\hat{\theta})$ and (\ref{EQ:CI_g_October_23_2022}) approach is 0.949. Similarly, the squared biases for $\hat{\Psi}_{\mathrm{BC}}(\hat{\theta})$ are virtually 0, while the squared biases of $\hat{\Psi}(\hat{\theta})$ are substantially larger.
Fig. \ref{fig:GAM_annual_temperature_anomalies_January_4_2023} demonstrates that when the bias is large the corresponding CI coverage rate is less than the nominal value due to an underestimation of the bias-squared term in MSE with the mean squared bias in (\ref{EQ:mse_hatpsi_f_approx}).
When the squared bias is small, the corresponding CI coverage rate is larger than the nominal value due to an overestimation of the bias-squared term in MSE.
Hence, the lack of accuracy in the estimation of the bias-squared term in MSE is an important reason for the unsatisfactory component-wise coverage of the CIs constructed with $\hat{\Psi}(\hat{\theta})$ and marginal MSE.

\FloatBarrier
\section{Discussion}\label{sec:discussion}

We investigated statistical inference for nonlinear mixed-effects models for the case when random effects (REs) were used as a pragmatic approach to estimate high dimensional fixed effect parameters (FEs).
This case includes GAMs and other nonparametric regression methods. If the FEs possess some intrinsic correlation, i.e., FEs that are close together spatially and/or temporally tend to have similar values, then modeling them as REs with multivariate correlation structure can be a suitable estimation procedure, and the conditional inference discussed in this paper is appropriate. This is also applicable to some implementations of mixed-effects models where the REs actually remain fixed during repeated samplings, and hence conceptually should be classified as FEs.	We proposed a novel procedure for less biased estimation of the REs, and improved confidence intervals. We demonstrated using simulation studies that our new methods provide substantially improved statistical inferences. 

For conditional confidence interval (CI) when there are REs not associated with data, we can have an outlook alternative to that in Sec. \ref{sec:conditional_CI_no_data_December_20_2022}. Let $\Psi_A$ be the subset of REs $\Psi$ associated with data $D$, and $\Psi_B$ be the subset not associated with data, which can be represented by a conditional independence statement 
\begin{linenomath*} 
	\begin{align}\label{EQ:conditional_indepence_April_11_2023}
		f(D|\Psi_A,\Psi_B,\theta)&=f(D|\Psi_A,\theta).
	\end{align} 
\end{linenomath*}
Therefore, all the conditional quantities such as mean and variance given $\Psi$ are actually conditioning on $\Psi_A$. The estimates of $\theta$ and $\Psi_A$ are based on the joint density $f(D|\Psi_A,\theta)f(\Psi_A|\theta)$, namely, $\theta$ are estimated by $\hat{\theta}$ maximizing the marginal likelihood $f(D|\theta)$ integrating out $\Psi_A$, and $\Psi_A$ are estimated using the posterior mode, or equivalently, the mode of the joint density $f(D|\Psi_A,\hat{\theta})f(\Psi_A|\hat{\theta})$ respecting $\Psi_A$. Since all the elements of $\Psi_A$ are associated with data, its bias-corrected estimator and CI can be constructed with (\ref{EQ:bias_corrected_December_20_2022})--(\ref{EQ:CI_Psi_October_23_2022}). This whole inferential procedure does not involve $\Psi_B$; that is, the estimation of $\Psi_B$ belongs to prediction of ``future", even though $\Psi_B$ can spatiotemporally interweave with $\Psi_A$ over the range of the latter. Mathematically, the predictor $\hat{\Psi}_B$ of $\Psi_B$ should be a function of the estimator $\hat{\Psi}_A$ of $\Psi_A$. If $\Psi=(\Psi_A^{\top},\Psi_B^{\top})^{\top}$ follows a multivariate normal (MVN) distribution, this function has an analytical form obtained as follows. Let the precision matrix of the RE density $f(\Psi_A, \Psi_B|\theta)$ be 
\begin{linenomath*} 
	\begin{align*}
		\Sigma^{-1}=\left( \begin{tabular}{cc}		
			$\Lambda_A$	& $\Lambda_{AB}$ \\		
			$\Lambda_{BA}$	&  $\Lambda_B$
		\end{tabular} \right),
	\end{align*} 
\end{linenomath*}
which has been partitioned according to the dimensions of $\Psi_A$ and $\Psi_B$. Prediction of $\Psi_B$ is obtained by maximizing the log-joint likelihood $\log f(D|\hat{\Psi}_A,\hat{\theta}) + \log f(\hat{\Psi}_A, \Psi_B|\hat{\theta})$ with respect to $\Psi_B$, which gives the normal equation $\Lambda_{BA}\hat{\Psi}_A + \Lambda_B \hat{\Psi}_B = 0$, and hence 
\begin{linenomath*} 
	\begin{align}\label{EQ:Psi_B_solution_April_11_2023}
		\hat{\Psi}_B &= -\Lambda_B^{-1}\Lambda_{BA}\hat{\Psi}_A.
	\end{align} 
\end{linenomath*}
Due to the conditional independence property (\ref{EQ:conditional_indepence_April_11_2023}), there is no inference conditional on $\Psi_B$. As a result, a bias-corrected estimator cannot be constructed for $\Psi_B$, and marginal inference has to be adopted for $\Psi_B$; that is, the marginal MSE (\ref{EQ:mse_hatpsi_f_approx}) is applied to set up CI for $\Psi_B$. The covariance between $\hat{\Psi}_B$ and the bias-corrected estimator for $\Psi_A$ conditional on $\Psi_A$ can be derived using (\ref{EQ:cov_hat_Psi_conditional_April_11_2023}) and (\ref{EQ:Psi_B_solution_April_11_2023}). In order to partition $\Psi$ into $\Psi_A$ and $\Psi_B$, we suggest to use the singular value decomposition of the Hessian matrix of $\log f(D|\Psi_A,\theta)$ with respect to $\Psi$. The components of $\Psi$ with singular values virtually equal to zero are $\Psi_B$ and the others are $\Psi_A$. In the random-walk example of Sec. \ref{sec:random_walk_missing}, where the last three time points had no data, the prediction inference for new REs were basically applied. The above analysis can be developed into a method to deal with REs without data. Our simulations indicated that this method performed close to the method proposed in Sec. \ref{sec:conditional_CI_no_data_December_20_2022}; hence, we only fully formulate the latter method in this paper.

%\begin{acknowledgements}
%\end{acknowledgements}

\bibliographystyle{spbasic}      % basic style, author-year citations
\bibliography{mybib_marine}   % name your BibTeX data base


\end{document}